# Pluto's Far Side

S.A. Stern
Southwest Research Institute

O.L. White
SETI Institute

P.J. McGovern
Lunar and Planetary Institute

J.T. Keane
California Institute of Technology

J.W. Conrad, C.J. Bierson
University of California, Santa Cruz

C.B. Olkin
Southwest Research Institute

P.M. Schenk
Lunar and Planetary Institute

J.M. Moore
NASA Ames Research Center

K.D. Runyon
Johns Hopkins University, Applied Physics Laboratory

and The New Horizons Team

# Abstract

The New Horizons spacecraft provided near-global observations of Pluto that far exceed the resolution of Earth-based datasets. Most Pluto New Horizons analysis hitherto has focused on Pluto's encounter hemisphere (i.e., the anti-Charon hemisphere containing Sputnik Planitia). In this work, we summarize and interpret data on Pluto's "far side" (i.e., the non-encounter hemisphere), providing the first integrated New Horizons overview of Pluto's far side terrains. We find strong evidence for widespread bladed deposits, evidence for an impact crater about as large as any on the "near side" hemisphere, evidence for complex lineations approximately antipodal to Sputnik Planitia that may be causally related, and evidence that the far side maculae are smaller and more structured than Pluto's encounter hemisphere maculae.

## Introduction

Before the 2015 exploration of Pluto by New Horizons (e.g., Stern et al. 2015, 2018 and references therein) none of Pluto's surface features were known except by crude (though heroically derived) albedo maps, with resolutions of 300-500 km obtainable from the Hubble Space Telescope (e.g., Buie et al. 1992, 1997, 2010) and Pluto-Charon mutual event techniques (e.g., Young & Binzel 1993, Young et al. 1999 2001). The flyby of Pluto by New Horizons revolutionized knowledge of Pluto in many ways, including via the moderate and high-resolution imaging of its near side (i.e., closest approach) hemisphere.

However, owing to the fact that Pluto is a slow rotator with a 6.3872-day period, a single spacecraft fast flyby like New Horizons could only observe one hemisphere of the planet closely. Nonetheless, the flyby approach was planned to return daily or more frequent panchromatic and color images from the spacecraft's LORRI (LOng Range Reconnaissance Imager; Cheng et al. 2008) and MVIC (Multispectral Visible Imaging Cameras) medium focal length (Reuter et al. 2008) imagers. Although the resolution of these images are factors of ~20-50× worse than the global mosaics of the near side (NS) hemisphere (see Figure 1), they still represent vast (i.e., 15-30×) resolution improvements over what was available on Pluto's so-called far side (FS) prior to the New Horizons flyby.

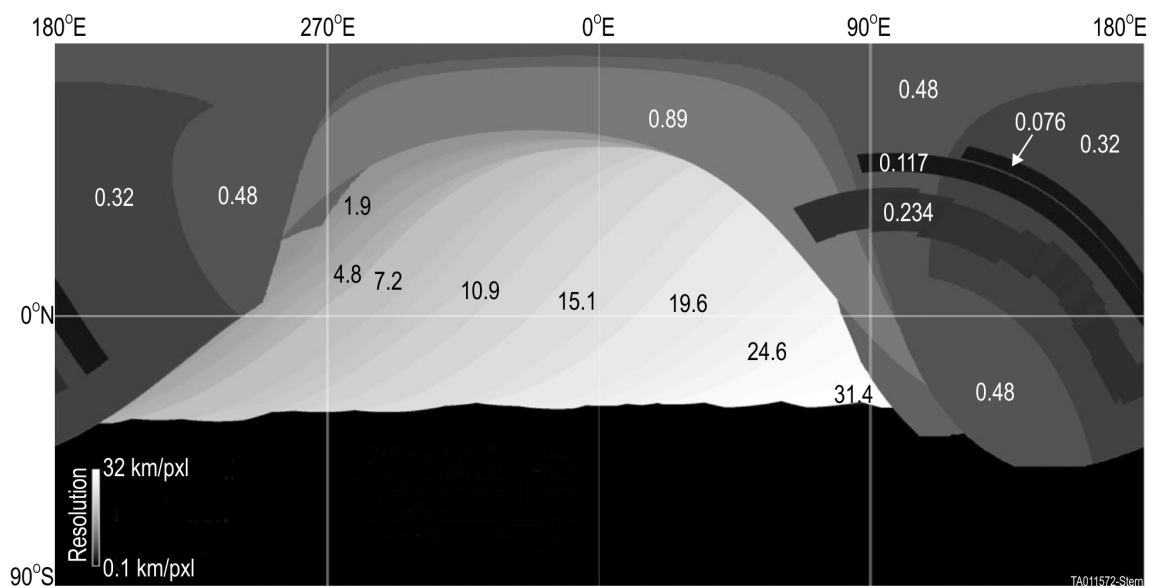

**Figure 1.** Global cylindrical map, centered at 0°E, illustrating the variation in pixel scale of the best New Horizons panchromatic imaging across Pluto. The black region in the southern hemisphere was not imaged by New Horizons as it was in winter darkness during the flyby.

Here we summarize the main far side results obtained with these imagers and describe the albedo, color, and geological interpretations that can be gleaned from existing inspections of these data. We also point to both specific needs for future work to learn even more about Pluto's FS and to the prospects for future FS studies in the coming decades.

## Far Side Maps

Here we present far side panchromatic and color mosaic products that initiate our discussion of the FS. Figure 2 contains a global panchromatic map, which has a characteristic FS resolution of 20 km/pixel. Because of the New Horizons approach vector, areas of the FS north of ~50° latitude were observed at ~0.3-1 km/pixel resolutions, but areas south of this were only imaged during the approach phase at FS resolutions of 5 to 30 km/pixel. This map has been modified from the official global map product described in Schenk et al. (2018) and archived in the PDS. Although positionally identical to published maps, the high resolution mapping coverage within the FS in Figure 2 has been extended laterally by cropping the images to a larger emission angle (lower values were used in the published product because of projection smear, but retained here in order to extend the highest resolution mapping into the FS as much as practical). In addition, the low resolution imaging portion of the map was reprocessed with data observed at emission angles >50° cropped. Although this means slightly lower resolution data were mapped at each location, this processing removes oblique projection distortions inherent in the originally published map.

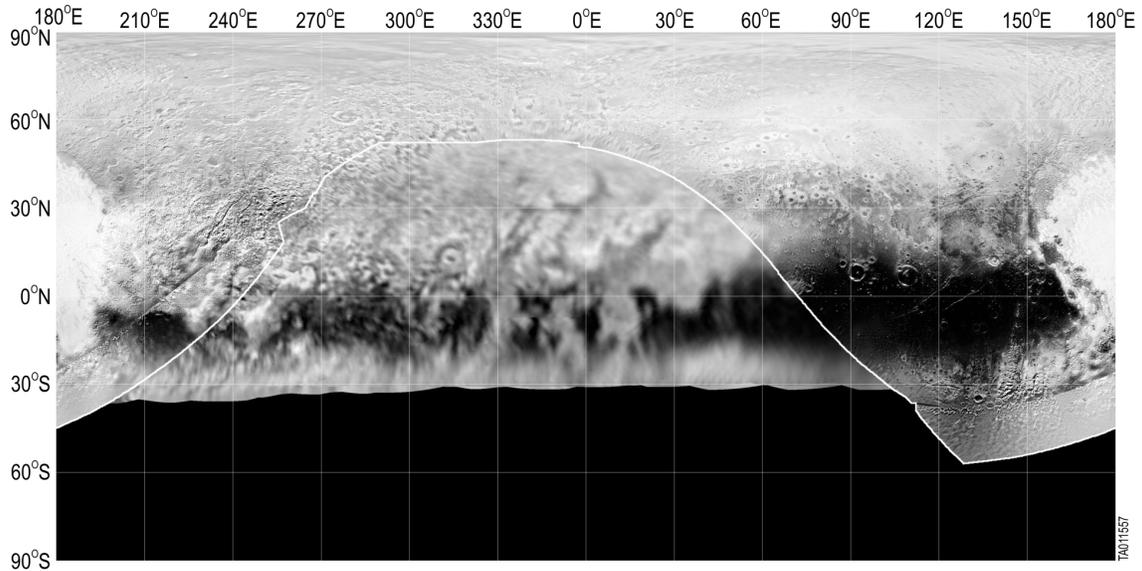

**Figure 2**. Global, panchromatic image mosaic of Pluto produced at 300 m/pixel. This cylindrical map projection is centered at 0°E on the far side. This figure's black areas below ~38°S were not illuminated during the New Horizons flyby. The area above the white line is covered by NS hemisphere imaging (resolution better than 1 km/pixel); the area below the white line is covered only by far side imaging (>1 km/pixel resolution).

Figure 3, adapted from Buratti et al. (2017), shows a global bond albedo map. Inspecting Figures 2 and 3 reveals that the FS is more uniformly structured in latitude (i.e., zonally structured), primarily due to the absence of the near side hemisphere's dominating albedo anomaly of Tombaugh Regio. It is also clear that the FS maculae are individually smaller (i.e., less extensive) and have more internal albedo variation than the maculae of the NS hemisphere.

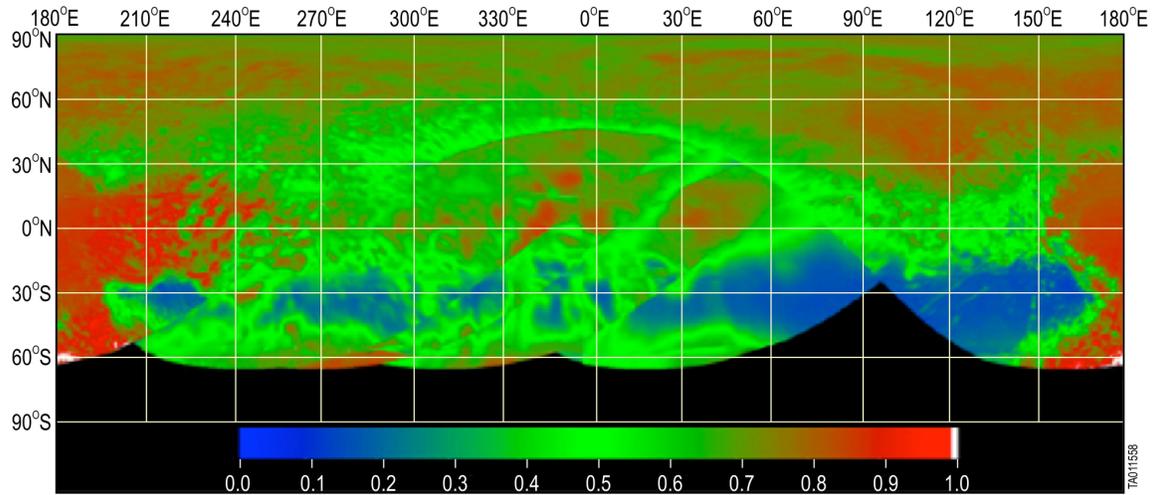

**Figure 3**. Published global Bond albedo map (adapted from Buratti et al 2017).

Figure 4 presents a global color map adapted from Schenk et al. (2018), using red, green and IR MVIC color filter imaging. These maps were processed to maximize color differences and show the relative redness and blueness (i.e., spectral slope) of each color unit. Similar conclusions to those drawn from Figures 1 and 2 can be drawn here as well. As described by Schenk et al. (2018) and Moore et al. (2018), relatively bluish units are concentrated along the equatorial band between 30°N and 30°S, and these have been ascribed by both authors as related to the presence of 'bladed terrains,' though there is somewhat more complex variegation in the distribution of blue color units. We discuss this more below in the context of FS bladed terrain deposits.

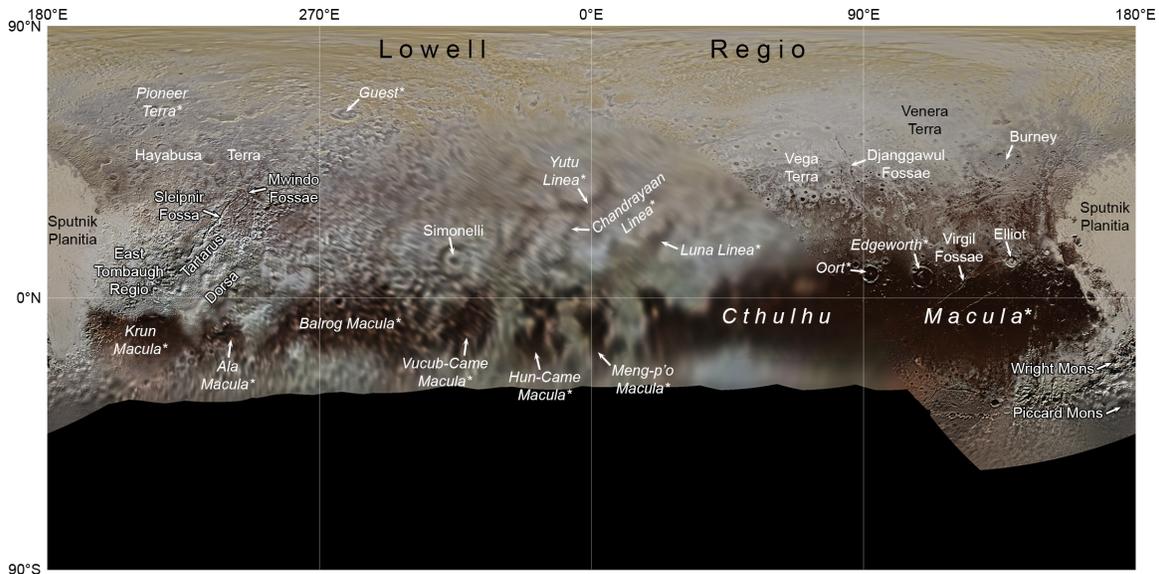

**Figure 4**. Global, cylindrically projected image mosaic of Pluto produced at 300 m/pixel overlain by MVIC color data produced at 650 m/pixel (from Schenk et al. 2018). The "blue," "red," and "$CH_4$" MVIC filters are used to provide the greatest color contrast. Labels indicate both formally and informally named features referenced in the text: formal names are shown in regular font, and presently informal names are italicized and asterisked.

## Geological Mapping

The great contrast in pixel scale between the near side and the far side means that contact definition and unit characterization for the FS must be based primarily on albedo variations seen in low-phase approach imaging, somewhat aided by interpretation from near side images. At the pixel scale of the FS imagery, which ranges from 2.2 km/pixel to 40.6 km/pixel, only surface features on a scale larger than ≈10 km (at the western extreme of the far side) and ≈200 km (at the eastern extreme) are well-resolved. Further, detailed surface textures that are apparent in near side hemisphere imaging of <1 km/pixel are invisible on the FS.

Given the inability to observe detailed surface textures directly on the far side, near side imaging that abuts the far side is an important anchor for FS mapping. In this way, unit contacts and large-scale structures that are easily defined in high resolution near side hemisphere coverage can be interpreted into the FS. Also, though the digital elevation model created for Pluto (Schenk et al. 2018) does not extend into the FS, FS topography can be gleaned from examination of the terminator in approach imaging of pixel scale ~5 km/pixel or better. And in addition, limb profiles, even though they are only sporadically distributed, allow us to assess topography on a scale of kilometers on the FS. These limb profiles represent an invaluable resource for unit definition to supplement visible and spectral imaging; see the text below on limb profiles for further details.

Our FS mapping follows standard US Geological Survey (USGS) mapping protocol (Skinner et al. 2018) and is shown in Figure 5. The following section describes each of the mapped units and explains our mapping rationale. Based on the continuations of albedo and color units in the base maps (Figures 1-4), we find most of the FS units to also be represented in the near side hemisphere to some degree, and so we include descriptions of their appearance in near side imaging as well where appropriate. We now discuss each FS unit type in turn.

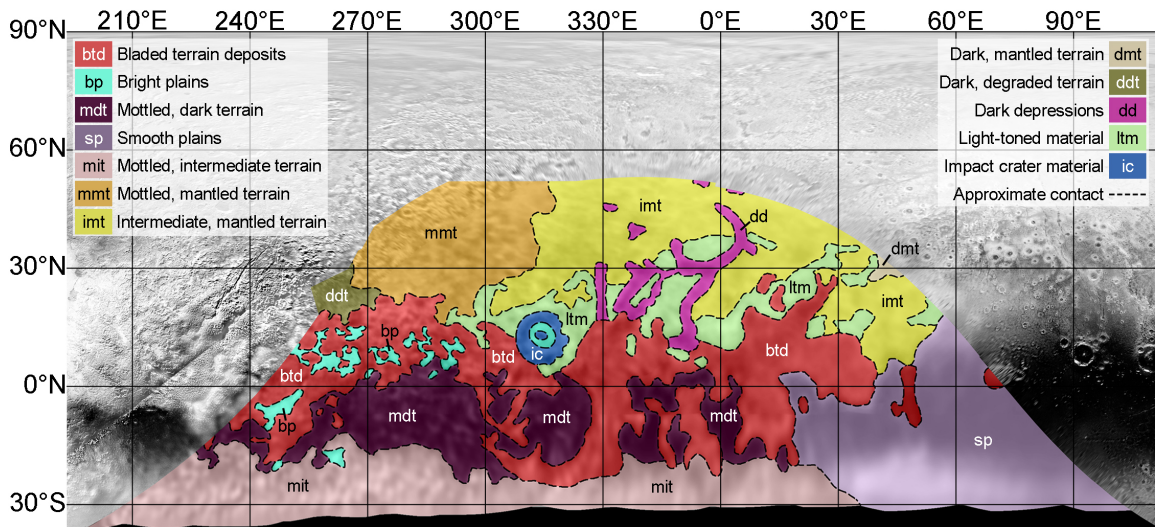

**Figure 5.** Geological map of Pluto's far side showing geological units identified by analysis of New Horizons imaging, spectral, and limb topography data. This map is overlain on photometrically equilibrated LORRI imaging ranging in pixel scale from 2.2 km/pixel (at the western boundary) to 40.6 km/pixel (at the eastern boundary), and which is surrounded by higher resolution, near side hemisphere imaging. Given the low resolution of imaging covering Pluto's far side, we treat all boundary contacts as approximate.

*Bladed terrain deposits (unit btd):* This unit was originally identified at the eastern boundary of the near side hemisphere (Moore et al. 2016, 2018; Moores et al. 2017), where it forms high-elevation (reaching >4.5 km above Pluto's mean radius), elongated, sub-parallel swells or plateaus (e.g., Tartarus Dorsa) that can display relief of more than 3 km above the plains to its north. The surface of these swells has a unique texture consisting of dense fields of typically sub-parallel sets of steep ridges that are characterized by sharp crests and divides, referred to as "bladed terrain" (Figure 6A). The blades are typically spaced 3 to 7 km crest-to-crest and exhibit relief of ~300 m from crest to base. Spectral data from New Horizons (Grundy et al. 2016; Protopapa et al. 2017; Schmitt et al. 2017) indicate that the bladed terrain deposits are composed primarily of methane ice (Moore et al. 2018). Moore et al. (2018) interpreted these massive deposits as having precipitated at high elevation and low latitude during an era in Pluto's history when climatic atmospheric conditions favored this formation scenario, with subsequent excursions in the climate causing the deposits to partially sublimate and hence erode into the blades seen today.

Bladed terrain deposits can be traced eastwards from the limit of their coverage in the haze-lit portion of the near side hemisphere. In low phase imaging, bladed terrain deposits appear as a low- to intermediate-albedo unit that contrasts strongly with the bright plains and dark maculae in this equatorial zone. In the western far side, the sub-parallel Tartarus Dorsa transition to a less ordered configuration, with the bladed terrain deposits surrounding expanses of bright plains (unit *bp*, see below), and also occurring as outcrops several tens of kilometers wide within these plains. East of 320°E, the bladed terrain deposits appear as irregular and angular formations hundreds of kilometers across that intersperse with dark equatorial maculae and separate these maculae from lighter toned terrain to the north.

In this far-eastern region of the far side, MVIC color observations and limb profiles are crucial to the identification of these formations as bladed terrain deposits. These correlate well to regions of high methane absorption as shown in Figure 14 of Moore et al. (2018), and limb profiles that cross this region (see below, Figure 8) indicate that they display relief of 2 to 4 km above adjacent terrain (i.e. similar to that of Tartarus Dorsa). The unit also shares a similar latitude range to Tartarus Dorsa, being located within latitudes 30° of Pluto's equator. Isolated outcrops of this unit in the far east of the far side are too small to be resolved in the spectral data, but are identified in limb topography. The easternmost outcrop of all, located at 70°E, 4°N, actually extends into the near side hemisphere. These observations indicate that Tartarus Dorsa forms the western extreme of a vast belt of bladed terrain deposits extending within an equatorial zone across >220° of longitude, primarily on the FS. The only major gap in this distribution is between longitudes 75° and 210°E.

*Bright plains (unit bp):* Like the bladed terrain deposits, this unit was also identified in mapping of the eastern near side hemisphere (Moore et al. 2018), where it appears as bright, generally smooth, but sometimes slightly hummocky, plains that occur on the floors of basins within East Tombaugh Regio and Tartarus Dorsa. This unit displays a lightly pitted texture, with pits reaching <1 km in diameter. It extends into the western FS, where it appears as bright terrain (with the highest albedo of any imaged far side unit) that forms irregular and angular expanses that occur interstitially to the bladed terrain deposits, and which can also embay outcrops of the deposits. The areal coverage of the unit decreases and becomes more sporadic with increasing longitude. The unit is interpreted as ponded volatile ices, specifically a solid solution of nitrogen ice and carbon monoxide with trace amounts of methane (Stern et al. 2015; McKinnon et al. 2016; Trowbridge et al. 2016) that have collected in depressions amongst the bladed terrain deposits ('1' labels in Figure 6A).

These ices form a consistent veneer across the whole landscape in East Tombaugh Regio and are seen to pond in depressions there, but the veneer dissipates at around 220°E, and their coverage in the western far side is limited to ponded deposits in such depressions. The main sequence of bright, volatile ice deposits terminates at ~290°E, but the floor of Simonelli crater (unit *ic*) at 315°E is also seen to display a high albedo, surrounding a dark central peak. Given that the ~250 km diameter Simonelli exists at the same latitude (12°N) as the well-resolved, 85 km diameter, central peak crater, Elliot, located in the far east of Cthulhu Macula, which displays bright deposits of volatile ices on its floor, it is reasonable to suspect that Simonelli's floor is also covered by such condensed ices, and we have mapped it as such. These observations are consistent with the west to east surficial compositional sequence described by Moore et al. (2018), from dominance by nitrogen ice closest to the low elevation Sputnik Planitia, to the increasing dominance of methane ice to the east, culminating in the high elevation bladed terrain deposits.

*Mottled, dark terrain (unit mdt):* A dark, equatorial band extends around nearly the entire circumference of Pluto, interrupted significantly only by Sputnik Planitia. This band is mostly contained within the permanent diurnal zone (Binzel et al. 2017) extending between latitudes ±13° of Pluto's equator. The cause of the low albedo can be interpreted to be a blanket of atmospheric haze particles that has accumulated on these surfaces (Moore et al. 2016; Gladstone et al. 2016; Grundy et al. 2016; 2018), which, once established as a consistent, dark mantling deposit, is kept warm enough by the diurnal Sun to prevent it from becoming a cold trap for condensation of bright, volatile ices (Binzel et al. 2017). An alternate hypothesis is that these terrains represent a substrate being exposed by the removal of seasonal volatile deposits in these regions. The bright methane deposits that are seen on north facing slopes of some crater walls in the vicinity of Pigafetta Montes in eastern Cthulhu Macula (Figure 13 in Moore et al. 2018) may represent such seasonal deposits. In this scenario, the wispy, mottled portions of *mdt* are the remnants of these bright, seasonal deposits.

Surface features underneath this blanket often appear sharp and well-defined in imaging ranging from 76 to 890 m/pixel, meaning that any dark blanket of material is not thick enough to mask or soften topographic relief on a scale of hundreds of meters to kilometers, and so the topographic signatures of individual geologic units are preserved. Grundy et al. (2018) estimate that unperturbed haze particle accumulation would coat the surface to a thickness of ~14 m over the age of the solar system, consistent with its masking effects being below the resolution of New Horizons imaging.

The FS portion of this band extends from Krun Macula at 225°E to Cthulhu Macula at 80°E. The dark mantling deposit is not itself a geological unit, but covers a great variety of different terrains. Near side hemisphere imaging shows that Krun and Cthulhu Maculae are very different geologically. Krun consists of a rough upland plateau with interconnected complexes of pits, troughs, and basins up to 3 km deep and 20 km wide (Howard et al. 2017; Moore et al. 2018), indicating that widespread surface collapse has affected this terrain ('2' labels in Figure 6B). Few recognizable impact craters exist there. Cthulhu consists of smooth (at a scale of kilometers), cratered plains that are crossed by tectonic belts. The dichotomy between Krun and Cthulhu indicates that there must be at least one geological transition within the maculae of the far side, and such a transition is identified at ~15°E, based on two criteria: (1) An expanse of bladed terrain deposits divides the band of maculae here, with terrain to the west generally having a more mottled appearance, whereas that to the east appears homogeneous by comparison (the differing texture is not an artifact due to degrading resolution with increasing longitude, as the difference is apparent in approach images with consistent resolution from west to east); and (2) In a limb profile that crosses both types of maculae (panel D in Figure 8), maculae to the west of 15°E are more elevated than those to the east.

The relatively high topographic relief of the mottled, dark terrain and its close spatial association to the bladed terrain deposits (whereby both are seen to embay occurrences of the other) suggest that a genetic relationship between the two may exist. The mottled texture of the unit could represent poorly resolved outcrops of lighter-toned bladed terrain protruding above the dark mantle that covers the terrain; alternatively, it may indicate a craggy and pitted texture akin to that of Krun Macula, and so represent an eastwards extension of this terrain type or the remnants of bright seasonal deposits of volatiles as mentioned above regarding the *mdt* unit.

*Mottled, intermediate terrain (unit mit):* The region south of the dark equatorial band on the far side consistently appeared only as a narrow strip along the limb in approach imaging, and so was only ever viewed at very high emission angles (>80°). As such, it appears smeared when re-projected in the global mosaic. Along with the low resolution of this imaging, this smearing further confounds characterization of this region. The unit has an overall intermediate albedo, but also displays albedo contrasts that form a mottled texture on a scale of tens of kilometers, a texture that is tentatively identified as far east as 30°E, where there is a transition to lighter, more homogeneous terrain that has been mapped as smooth plains (unit *sp*). In approach imaging, this intermediate albedo, mottled texture is also seen to characterize terrain south of Krun Macula within the near side hemisphere, and in 320 m/pixel haze-lit imaging it is seen to have a rubbly, knobby texture on a scale of several kilometers (Figure 6B). There are no recognizable impact craters here, although there are rimless depressions (termed cavi) that reach tens of kilometers across with dark deposits on their floors ('3' labels in Figure 6B). The presence of the cavi and the apparently youthful surface of this terrain suggest that it may be related to the tentatively cryovolcanic edifices Wright and Piccard Montes to the west (Moore et al. 2016; Schenk et al. 2018; Stern et al. 2018), although there are no constructional landforms. This rubbly terrain likely continues eastwards into the far side to form at least the westernmost portion of the mottled, intermediate terrain, but the low quality of the imaging here prevents any differentiation of this unit between 225°E and 30°E. There is the possibility that bladed terrain deposits extend further south into this unit than has been mapped in Figure 5, but neither visible imaging nor spectral data (i.e. methane absorption) can confirm this conclusion.

*Smooth plains (unit sp):* At the western edge of the near side hemisphere, terrain south of ~30°N is occupied by cratered plains (Figure 6C) that are crossed by tectonic systems including Djanggawul and Virgil Fossae. These plains display a dark cover of accumulated haze particles between ~13°N and ~23°S (Cthulhu Macula), and as noted previously, this cover has no discernable effect on the topographic relief of underlying landforms at the pixel scale of the imaging here (890 m/pixel). The dark equatorial plains and intermediate albedo plains on either side of them to the north and south can therefore be regarded as a single unit. On the far side, Cthulhu Macula appears as an expanse of homogeneous, dark terrain. The contact separating the northern, intermediate albedo plains from terrain to the north (the brighter intermediate mantled terrain and light-toned material, units *imt* and *ltm*) can be traced southwestwards, tapering until it meets the eastern limit of unit *btd* at ~40°E. Intermediate albedo plains are not identified west of 30°E, where the unit is only expressed as dark terrain bounded to the north by the bladed terrain deposits, and to the south by the mottled, intermediate terrain. An expanse of bladed terrain deposits separates the smooth plains from the irregular and angular maculae located between 330°E and 20°E to the west, which we interpret to be occurrences of the mottled dark terrain. Limb profiles confirm that bladed terrain deposits are elevated above the smooth plains by a few kilometers (Figure 8), with the former interpreted as being superimposed upon the latter. This can be observed directly just short of the limb in near side hemisphere imaging ('4' label in Figure 6C). The presence of scattered, intermediate albedo bladed terrain deposits within these cratered, darkened plains suggests that the bladed terrain deposits colonized high terrain here early in Pluto's history, before a dark mantle accumulated, and have experienced sufficient resurfacing via sublimation erosion since then to avoid subsequent mantling by the haze particles (in contrast to the plains surrounding them).

*Dark, degraded terrain (unit ddt):* In 890 m/pixel near side hemisphere imaging obtained at high phase (solar incidence >80°), terrain in the vicinity of the radial tectonic features Mwindo Fossae displays flat plains interspersed with jagged and degraded hills that have a somewhat arcuate morphology (Moore et al. 2018) ('5' labels in Figure 6D). In low phase approach imaging, this same terrain appears dark relative to surrounding terrain, and extends into the far side to the east of Mwindo Fossae. Geological mapping in Moore et al. (2018) indicates that the arcuate terrain underlies the bladed terrain deposits, and it was suggested that a genetic relationship exists between the two units, with the arcuate terrain possibly being marginal deposits left behind after retreat of more extensive bladed terrain deposits. The dark, degraded terrain may therefore represent ancient water ice crust that was previously covered by the bladed terrain deposits prior to their recession, and which could have extended north as far as the southern limit of the mantled terrains (30°N-40°N).

*Mottled, mantled terrain (unit mmt):* Across all longitudes, the northern latitudes of Pluto differ from the equatorial regions in that they display a conspicuously homogeneous and intermediate albedo compared to the strong albedo contrasts exhibited by the equatorial regions, as manifested most obviously in the dark maculae and the bright Tombaugh Regio. This albedo difference has been attributed to the fact that the high latitudes have experienced extreme insolation and seasonal cycles across Pluto's history, circumstances that should result in considerable and ongoing volatile exchange in response to such cycles, and which should be recorded to some extent in Arctic landscapes (Binzel et al. 2017; Earle et al. 2018). This response is in contrast to the equatorial regions, where temperature modeling calculations show that consistent diurnal variations are effective in long-term preservation of whatever material is "seeded" there (Binzel et al. 2017; Earle et al. 2017); here, evolution of landscapes subject to modification by exogenic processes would instead reflect secular, irreversible changes in Pluto's climate.

The area to the northeast of Sputnik Planitia is characterized by rounded terrain that is smooth-textured on a scale of a few kilometers, mapped as "smooth uplands" by Howard et al. (2017). This morphology is diagnostic of accrescence, whereby deposition of a thick mantle has occurred uniformly over a regional surface, causing projections to become rounded and inward facing, and valleys to become sharply indented. East of ~260°E, this mantled terrain takes on a mottled appearance, with the lighter-toned mantling deposit forming irregular, lobate, flat-topped plateaus that are separated by lower albedo depressions, their lower elevation resolved by the topographic mapping of Schenk et al. (2018) (Figure 6E). This terrain continues eastward into the far side, where it is mapped as mottled, mantled terrain (unit *mmt*), as far as 310 to 320°E. We interpret the lobate appearance of the plateaus to also be an accrescence texture, implying that this unit is where mantling of the surface has occurred to an incomplete degree, leaving gaps where the darker substrate remains exposed. At 61°N, 278°E, the mantle partly in-fills a 115 km diameter impact crater ('6' label in Figure 6E). Howard et al. (2017) mapped an expanse of low albedo "eroded, smooth plains" within Hayabusa Terra to the south of the brighter, thickly mantled smooth uplands, and the mottled, mantled terrain, which appears to display characteristics of both these terrains and may record a history of alternating deposition and erosion (the latter occurring at least in part via sublimation) of the mantling material.

*Intermediate, mantled terrain (unit imt):* Tracing the northern boundary between the near side hemisphere and the far side, the mottled, mantled terrain transitions at ~310-320°E to a terrain where the dark-floored depressions are far fewer, not connected, and sporadically distributed ('7' and '8' labels in Figure 6F). This appears to indicate that coverage by the mantling deposit is generally more consistent and complete here than in the mottled, mantled terrain. As such, FS terrain at this location has been mapped as intermediate, mantled terrain (unit *imt*). Given the very low resolution of the far side imaging (>15 km/pixel) in its central and eastern regions, we have designated most of the intermediate albedo terrain to the east as belonging to this unit, as no contacts can be identified within it, even tentatively.

Whereas the unit appears as a relatively consistent mantling deposit where it occurs at the boundary between the near side hemisphere and the far side, more generally we regard this unit as representing variably mantled terrains across the northern latitudes (i.e. mostly north of ~15°N) that nevertheless show a fairly consistent intermediate albedo in available imaging.

We also note that the northwest of the near side hemisphere displays a mantling deposit ('9' labels in Figure 6F) that has a higher albedo than the intermediate mantled terrain, it is tectonized and partly eroded, and it also shows impact craters with generally sharply defined rims and dark floors. This represents an extension of the "fretted terrain" described by Moore et al. (2016). Due to the particularly low resolution of far side imaging in this area, and the subtle difference in shading between this unit and the intermediate mantled terrain, it is difficult to determine if this bright mantling deposit extends into the far side at all, but we have not mapped it as doing so. Within the near side hemisphere, these two mantling deposits appear to make contact in the region around 62°N, 357°E, where the bright, mantled terrain lines the rims of the easternmost dark pits of the intermediate mantled terrain ('8' labels in Figure 6F).

*Dark, mantled terrain (unit dmt):* Within the western near side hemisphere, in Vega Terra, there is a mid-latitude, generally low albedo zone that has a mottled appearance on a scale of a few kilometers (in 232 m/pixel imaging). This zone seems to be where overlying, higher albedo material (possibly the bright, mantling material of the bright, mantled terrain to the north, or material of the smooth plains to the south) is undergoing degradation and removal, leaving darker material exposed underneath, although some of the dark material may be haze particles that have settled in depressions. The mottled texture continues eastwards as a mid-latitude band as far as Sputnik Planitia. Vega Terra also features the prominent "bright-halo craters", which display conspicuously bright inner and outer walls, and dark floors (Moore et al. 2016) (Figure 6G). To the west, this terrain can be discerned on the far side as an especially low albedo promontory (the dark, mantled terrain, unit *dmt*) extending into the intermediate, mantled terrain. It does not appear to extend very far into the far side (<150 km), although the very low resolution of the imaging here makes its true extent difficult to determine.

*Light-toned material (unit ltm):* In the FS low- to mid-latitudes there is a roughly defined band of light-toned material (unit *ltm*) that is brighter than units contacting it (specifically the bladed terrain deposits, intermediate, mantled terrain, and dark depressions), but is also not as bright as the volatile ice of the bright plains. This is the most tenuously defined far side unit. It borders our mapping area at one location in the far east, and we consider a particularly bright region here (label '10' in Figure 6G) to be what is perhaps its only representation in the near side hemisphere, although the imaging here is very oblique. It bounds the dark depressions (unit *dd*), relative to which it is elevated, and also the bladed terrain deposits (unit *btd*), relative to which it is depressed. The light-toned material may indicate precipitation of methane ice, e.g. like that observed at Pigafetta Montes or the bright halo craters (Moore et al. 2016, 2018). In this case it would not be a true geological unit, but rather a thin, discontinuous veneer that is covering the mottled, mantled terrain and the intermediate, mantled terrain. In addition, whereas the unit is not obviously sequestered within large depressions like the bright plains-forming material (unit *bp*), it cannot be discounted that bright volatile ice that has condensed within depressions that are too small to be resolved by the imaging in this part of the far side (>10 km/pixel) may also contribute at least in part to the high albedo of this unit.

*Dark depressions (unit dd):* The dark depressions cluster in a 40° wide zone of longitude that is centered at ~350°E, and extend north from the bladed terrain deposits. Here, they appear as lineations that reach hundreds of kilometers long and tens of kilometers wide. The morphologies of these lineations range from linear to sinuous to crescent-shaped. A single, high phase, high resolution (800 m/pixel) limb profile crosses these lineations (Figure 8B), confirming that they are depressed and not elevated (Figure 6H). This unit also includes a handful of northern, isolated, dark patches tens of kilometers across that are separated from the main cluster of lineations. These are interpreted to be larger versions of the dark pits seen to the north in the near side hemisphere ('7' and '8' labels in Figure 6F), a few of which border the far side. We interpret all of these depressions to be where the mantling material of the intermediate, mantled terrain has been removed, revealing a darker substrate. Within the near side hemisphere, a dark substrate is also exposed between the lobate plateaus of the mottled, mantled plains (Figure 6E), on the floors of impact craters within the bright, mantled terrain (Figure 6F), and in the dark, mantled terrain (Figure 6G).

Different mechanisms may have acted to remove the overlying mantle depending on location. If the isolated, far-northern patches are larger versions of the dark-floored pits seen within the intermediate, mantled terrain in the near side hemisphere, then these depressions may have formed by sublimation erosion of the mantle, as is hypothesized for the 'eroded smooth plains' of Hayabusa Terra (Howard et al. 2017), the mottled, mantled plains, and the dark, mantled plains. The concentration of the lineations between ~330°E and 10°E places them neatly in line with the NNE-SSW-aligned ridge-trough system that spans the entire near side hemisphere (Schenk et al. 2018), and if they are part of the system, then they may be where fracturing of Pluto's crust after deposition of the mantle has exposed the dark substrate. The oblique angle that some of the lineations make to the trend of this system, however, suggests that not all of them originate due to the same tectonic forces that produced it. The appearance of the dark crescent feature that is centered at 38°N, 357°E is somewhat reminiscent of the smaller, partly mantled crater at 61°N, 278°E ('6' label in Figure 6E), and may potentially represent a 350 to 400 km diameter, dark-floored impact crater that has been mostly in-filled with the mantling deposit of the intermediate, mantled terrain, with the exception of a narrow margin along its southern and eastern walls. Alternatively, the lineations, in addition to the irregular and angular bladed terrain deposits and maculae that occur immediately to the south of them, may instead represent a >1000 km diameter zone of crustal disruption that is located antipodal to the location of the Sputnik basin-forming impact. Similar zones of chaotic terrain that occur antipodal to large impact basins are seen elsewhere in the solar system, notably on Mercury (e.g., Schultz & Gault 1975). Such disruption (and accordingly the Sputnik impact) would have necessarily occurred after the deposition of the mantle.

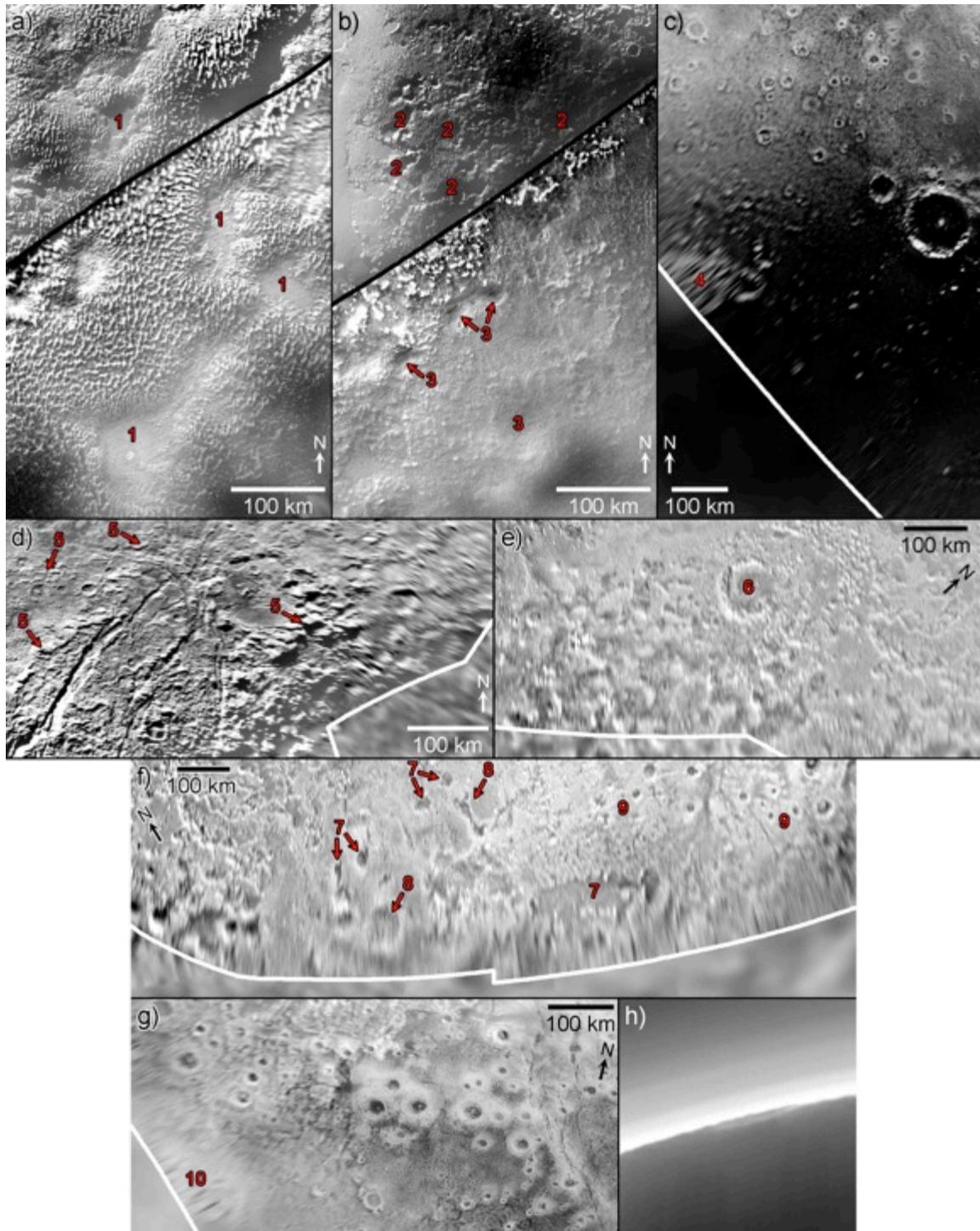

**Figure 6.** Various terrains within the near side hemisphere that border, and which are interpreted to form geological units within Pluto's FS;

black lines indicate the boundary between the directly illuminated portion of the near side hemisphere, and the haze-lit portion. Thick white lines indicate the boundary between the near side hemisphere and far side. (a) Bladed terrain deposits of Tartarus Dorsa (unit *btd*). '1' labels indicate deposits of bright, volatile ices (unit *bp*) ponded on the floors of depressions amongst the dorsa. Image is centered at 4.1°N, 228.8°E. (b) Dark, eroded material of Krun Macula (above black line), which may be representative of unit *mdt*, and intermediate albedo, knobby terrain (below black line) that may be representative of unit *mit* (at least its western region). '2' labels indicate large collapse pits within Krun Macula, and '3' labels indicate rimless, dark-floored depressions (or 'cavi'). This image is centered at 17.8°S, 198.4°E. (c) Smooth, cratered plains (unit *sp*), with the plains in the lower half of the image displaying a veneer of accumulated haze particles (Cthulhu Macula). '4' label indicates an expanse of bladed terrain deposits superposing the smooth plains. Image is centered at 4.5°N, 82.0°E. (d) Dark, degraded terrain (unit *ddt*) in the vicinity of Mwindo Fossae. '5' labels highlight occurrences of jagged, hilly terrain with a distinctly arcuate morphology. This image is centered at 31.0°N, 250.1°E. (e) Mottled terrain interpreted as a thick, light-toned mantling deposit partly covering a darker substrate (unit *mmt*). '6' label indicates an impact crater with the mantle partly covering its floor. This image is centered at 57.5°N, 282.5°E. (f) The intermediate, mantled terrain (unit *imt*). '7' labels indicate isolated, dark floored pits that are likely smaller versions of the northern dark depressions seen in the far side (unit *dd*). This area marks a transition to a brighter, tectonized/mantled terrain ('9' labels). '8' labels indicate pits with high albedo material characteristic of this bright, mantled terrain occurring around their rims. Albedo contrast has been stretched in this image. This image is centered at 59.2°N, 4.5°E. (g) Mottled terrain interpreted as a light-toned mantling deposit that is undergoing erosion to reveal a darker substrate (unit *dmt*). Bright-rimmed ("halo") craters occur frequently. '10' label indicates an especially bright region that may represent the light-toned material (unit *ltm*). This image is centered at 31.6°N, 69.8°E. (h) The limb of a high phase (170°), 800 m/pixel LORRI observation that crosses a lineation of the dark depressions (unit *dd*), located at ~31.5°N, 3.5°E. The near-side rim of the lineation appears as a ~70 km wide notch in the limb, with the far-side wall of the lineation appearing brighter due to intervening haze in Pluto's atmosphere.

*Simonelli impact crater material (unit ic):* The only unambiguous impact feature that has been identified on the FS is the ~250 km diameter featured named Simonelli, located at 12°N, 315°E; see Figure 7. Its status as a central peak crater is confirmed by low Sun terminator imaging, where its walls and central peak appear sunlit and there is clear evidence of a raised annual rim (Figure 7A). At higher solar illumination, its walls and central peak appear dark, with a bright annulus separating them (Figure 7B). This annulus is interpreted to be deposits of volatile ices that have condensed on its floor, similarly to the somewhat smaller, 85 km wide Elliot crater in the near side hemisphere (Figure 7C). Bordered by bladed terrain deposits to its south and located within an area that shows high methane absorption (Moore et al. 2018), this impact may have occurred into the bladed terrain deposits or it may have been modified by the subsequent deposition of bladed terrain materials.

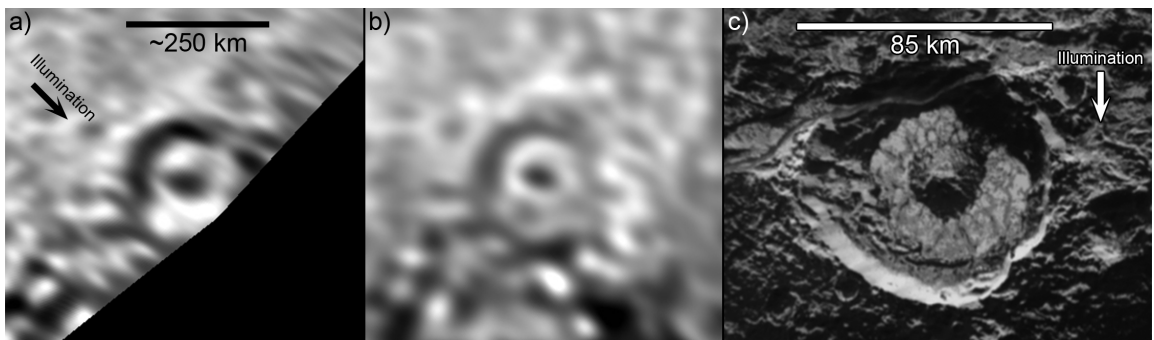

**Figure 7**. (a) Simonelli crater seen at nearly 90° phase in 9.3 km/pixel imaging, re-projected to cylindrical projection; illumination is from upper left. (b) Simonelli crater seen at nearly 0° phase angle in 16.9 km/pixel imaging, re-projected to cylindrical projection. (c) Elliot crater seen in 320 m/pixel imaging, with bright volatile ices condensed on its floor and surrounding its central peak.

## Topographic Knowledge

**Stereogrammetry and Photoclinometry.** Stereogrammetric mapping of FS topography was successful in areas north of ~50° latitude (Schenk et al. 2018). However, all attempts to derive coherent stereogrammetric Digital Elevation Model (DEM) data product southward of this in the FS were unsuccessful due to a combination of low parallax, low resolution and low inherent relief of these areas (likely <6 km; Schenk et al. 2018).

Photoclinometry cannot be reliably used in the low-resolution part of the FS mapping coverage due to severe albedo contrasts and the lack of significant resolvable shadowing. In spite of this limitation, examination of the terminator regions of the approach images reveals no deep shadows or strong highlights that would indicate large variations in topography. Thus, we conclude that no high "Olympus Mons" or deep "Valles Marineris" type features are revealed there.

**Limb Profiles.** The lack of high-resolution images of the FS of Pluto and the lack stereo topography except in the FS north require the use of alternative methods to derive topographic information. Topographic profiles for these regions can still be obtained by careful measurements of the limb in both close encounter and post-encounter backlit FS images. These two types of image illumination require slightly different techniques for accurately determining limb positions. For standard images (viewing the day side of Pluto) we use Method A of Nimmo et al. (2017). This method scans each row and column in the image away from the body center. In each scan the limb is taken to be the point where the brightness is half way between the face and background value; see Nimmo et al. (2017) for details.

In backlit images, the disk of Pluto and the background are dark to within the noise of the camera. There is, however, some forward scattered light from atmospheric hazes, which is brightest near Pluto and fades farther away from the disk. The sharp contrast between the dark disk and the bright haze is captured very clearly by taking the gradient of the image. Therefore, in backlit images, each row and column of the image is scanned away from the body center, and the limb is taken to be the location of the maximum gradient. For the backlit MVIC scan 0299181303, low signal to noise ratios prevented this gradient approach from working consistently. For this image, when scanning each row and column the limb was taken to be the point when the pixel brightness exceeded a set threshold. For all methods, spurious limb picks caused by cosmic ray hits or other artifacts are manually removed.

We have used limb profiles extracted from four New Horizons observations in our geological mapping and analysis, shown alongside their ground tracks in Figure 8. These include two LORRI observations (one front lit and one back lit) and two MVIC observations (both back lit). These each have a pixel scale of <1 km/pixel, which is sufficient to resolve FS topographic relief. The profiles have been processed via linear de-trending, which renders short wavelength topography to be accurate to within 5% error. These profiles are annotated with the locations of units mapped in Figure 5, and supplement the profile shown in Schenk et al. (2018).

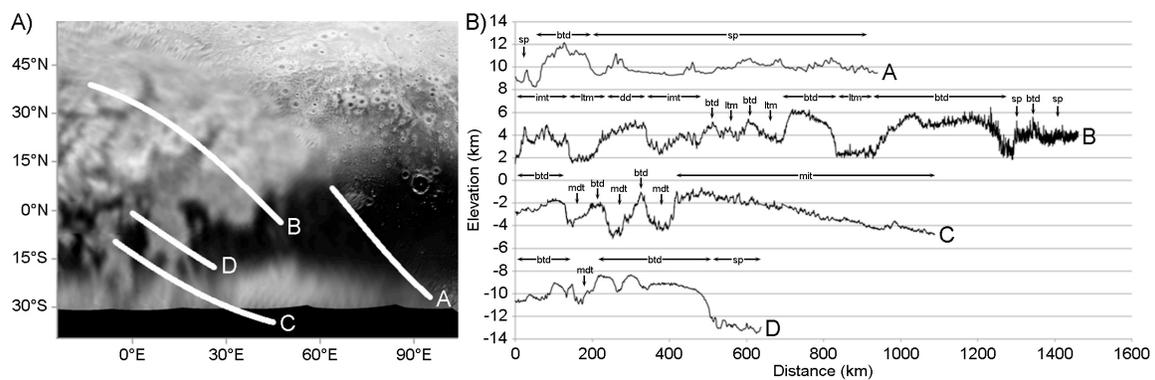

**Figure 8.** Panel A: Ground tracks of the four limb profiles shown in (Panel B). A is extracted from a single 889 m/pixel LORRI frame (0299168423). B is extracted from a mosaic of three 805m/pixel LORRI frames (0299192233, 0299192286, and 0299192339). C is extracted from a 346 m/pixel MVIC scan (0299181303). D is extracted from a 431 m/pixel MVIC scan (0299181722). Panel B: Profiles A-D, annotated with the locations of mapped units in Figure 5. Note that unit *dd* in profile B appears high as the depressed topographic signature of the lineation that the profile crosses here is being masked by background topography (see Figure 6h).

**Possible Tectonic and Sputnik-Antipodal Effects on the Far Side.** Tectonic trends of hemispheric extent identified on the well resolved near side hemisphere of Pluto cause us to consider the potential expression of such activity on the less well resolved far side. This is further motivated because in general it is known that basin-forming impacts can have major terrain-altering effects at locations antipodal to the point of impact, both from convergence of ejecta (Wieczorek &

Zuber 2001) and also due to focusing of shockwaves traveling around the solid planet (e.g., Schultz & Crawford 2011).

One of the most dramatic physiographic features on Pluto's near side hemisphere, discovered in the topographic mapping of Schenk et al. (2018), is a complex, eroded, fragmentary band of graben, troughs, ridges, plateaus, tilted blocks, and elongated depressions (or 'cavi') that extends at least 3200 km from the north pole to the limit of coverage at ~45°S, and which is ~300 to 400 km wide. Designated by Schenk et al. (2018) as the "great north-south Ridge-Trough System" (RTS); this NNE-SSW-trending band crosses the equator at 150°E. Its configuration is suggestive of an origin stemming from a planetary-scale loading mechanism.

The landforms that compose RTS appear to derive from extensional tectonism. The most plausible scenario to generate a wide band of terrain with the required orientation of extension is equatorial thickening of the water ice shell lithosphere of Pluto (McGovern et al. 2019). Under this hypothesis, the RTS represents a paleo-equator. Since the RTS now trends NNE-SSW, it would imply a significant amount of reorientation (i.e., true polar wander) of Pluto. True polar wander has been hypothesized for Pluto based on the location of the Sputnik Planitia glacier and Pluto's other global scale tectonic patterns (Keane et al. 2016; Nimmo et al. 2016). However, the true polar wander scenario in these papers does not put the RTS at the paleo-equator. A possible explanation is that the RTS predates the formation of Sputnik Planitia, and Sputnik Planitia-derived tectonics. It is curious to note that the orientation of the RTS would be consistent with a paleo-equator if Pluto reoriented about the minimum principal axis (the tidal axis connecting Pluto and Charon), and swapping the intermediate and maximum principal axis of inertia. This style of reorientation may naturally occur if Pluto's rotation and volatile loading cycles are coupled (e.g., Rubincam 2003).

Schenk et al. (2018) identified several dark-floored depressions in the well-resolved polar region of Pluto's far side, and suggested that other dark, irregularly shaped features in the poorly-resolved portion of the far side may also be depressed, possibly indicating that RTS continues well into the far side but with a different, less linear character. We have mapped these dark features as unit *dd*; limb topography confirms that they are depressions (Figure 6h).

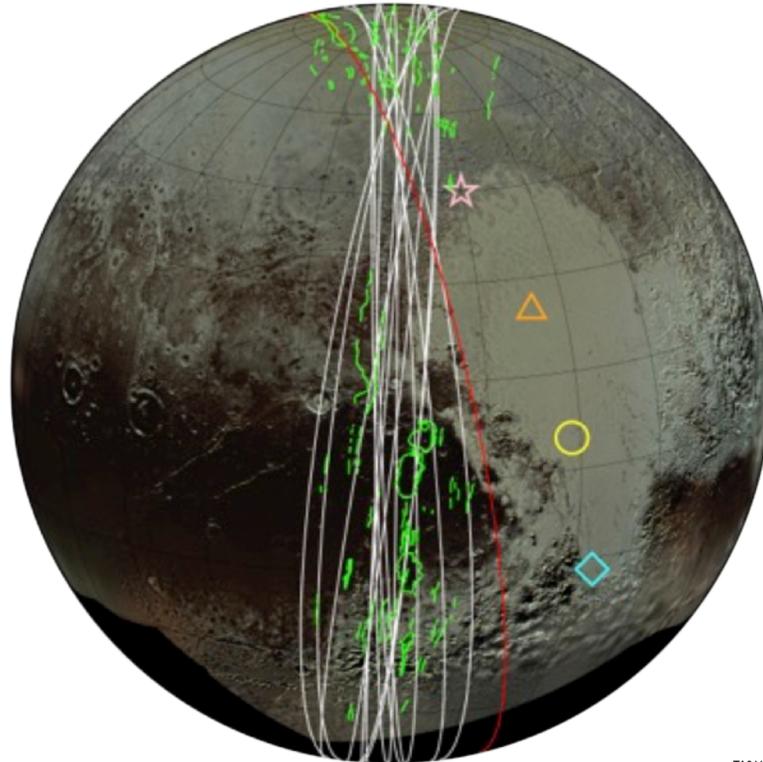

**Figure 9.** Spherical orthographic projection of the global mosaic of Pluto (Schenk et al. 2018) focusing on the near side hemisphere. Center of projection is 150°E, 15°N. Tectonic lineations are mapped with various colors. The green lineations indicate the tectonic system dubbed the "great north-south ridge-trough system", or RTS (Schenk et al. 2018). The white lines represent great-circle paths fit to individual lineation segments of RTS. The red line represents a great-circle path fit to the divergent northwestern segments of RTS. The star, triangle, circle, and diamond symbols represent a potential first contact point for an impact incidence from the NNW, the center of the deep portion of the Sputnik basin, a southern extension of SP basin-filling materials, and the potential first-contact point for an impact incidence from the SSE, respectively.

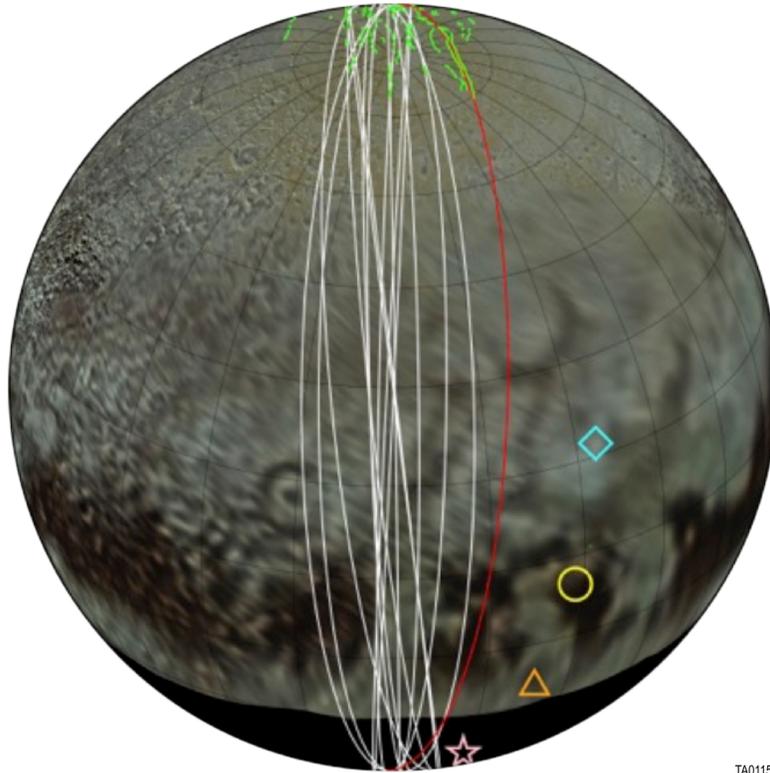

**Figure 10.** Global mosaic, as in Figure 9, but focusing on Pluto's far side. Center of projection 330°E, 30°N. Green lineations at the top of the figure represent the northernmost elements of the RTS. The star, triangle, circle and diamond symbols indicate the antipodes of the corresponding Sputnik Planitia basin-related locations in Figure 9.

Schenk et al. (2018) also noted that limb profiles in medium-resolution departure images indicate an unusually rugged topography near 320–340°W at ~30°N, which can be interpreted as an extension of the northern section of the system. Accordingly, our great-circle extrapolations of the trends of RTS's faults to Pluto's far side (Figure 10) show a concentration oriented along the 330°E meridian, antipodal to the 150°E alignment of RT in the near side hemisphere (Figure 9). This concentration overlaps with the westernmost lineation of unit *dd*, aligned almost exactly along this meridian (Figure 5), and also overlaps the NNE-SSW-oriented lineations of unit *dd* between 335° and 345°E. However, occurrences of unit *dd* lying further eastward, including the large crescent-shaped feature, do not line up convincingly with RTS great-circle pathways, except perhaps those fit to a group of the northernmost long segments of the RTS. The fractures of RTS splay in the north polar region, and so this group displays orientations significantly different from those of fractures in lower-latitude segments of RTS. The red line in Figure 10 shows the great circle of one of these polar fractures. Such great circle trends cross the westernmost part of the crescent and closely parallel the NNE-SSW oriented segment of unit *dd* emanating from the southernmost part of the crescent. We consider it significant that these uniquely oriented polar RTS segments are both closest to, and the only segments with trend orientations encompassing the easternmost components of unit *dd*. This combination points to a common regional-to-hemispheric influence on the expression of both RT and unit *dd* on Pluto's FS.

The RTS trend most likely continues to Pluto's FS, and may express itself in part as the large-scale, roughly north-south-aligned, depressed lineations of unit *dd*. Imaging limitations prevent us from determining the precise nature of these depressions, e.g., whether they are graben (like those seen in the polar portion of the RTS) or elongate cavi (like those seen in the equatorial portion of the RTS). Further south, we also consider the possibility that the north-south-aligned fabric of the RTS could also account for the narrow, angular, and roughly north-south-aligned segments of the bladed terrain deposits (unit *btd*) and mottled, dark terrain (unit *mdt*) between 340°E and 10°E.

Departures from the expected focus along the 330°E meridian and an essentially north-south alignment may be due to regional-to-hemispheric-scale variations in the structure of Pluto's ice shell and the stress field variations resulting from them. At a smaller scale, features like impact basins and volcanic edifices can account for the annular nature of mapped structures (such as the crescent feature), corresponding to impact basin rims, flexural moats of central volcanic edifices, or annular features termed "coronae", seen on Venus (attributed to volcano-tectonism or mantle upwelling; Stofan et al. 1997) and on the Uranian moon Miranda, which are attributed to diapirism in the icy crust (e.g., Pappalardo et al. 1997) or convection-driven resurfacing (Hammond & Barr 2014).

Pluto's smaller, northern, and more isolated components of unit *dd* (which are likely analogous to the irregular depressions indicated by labels '7' and '8' in Figure 6f) appear less likely to be tectonic in origin, and may be caused by degradation of the mantling material that covers the higher latitudes, perhaps by sublimation erosion. A continuation of the RTS trend to the FS would make the combined system a planet-encircling feature, and would strongly imply that a planetary-scale phenomenon is responsible for its formation. Because these particular patches of *mdt* are i) lower than the surrounding btd; and ii) elongate and broken up, in strong contrast to the more continuous or bulbous *mdt* units to the west, it is possible that the narrow and roughly north-south aligned segments of *mdt* between about 340° E and 10° E are southerly continuations of the *dd*/RTS trends.

Another potential influence on the geology of Pluto's FS is the antipodal response of a catastrophic event occurring in the near side hemisphere. The enormous depression that contains the massive nitrogen ice deposit of Sputnik Planitia has been interpreted to be an impact basin (Stern et al. 2015; Moore et al. 2016; McKinnon et al. 2016; Johnson et al. 2016; Keane et al. 2016; Nimmo et al. 2016; Schenk et al. 2018), in which case its elongated shape is suggestive of excavation by an oblique impact (McKinnon et al. 2016). Elliptical and elongated craters are created by low-incidence-angle impacts (Elbeshausen et al. 2013); examples include the South Pole–Aitken impact basin on the Moon (e.g., Garrick-Bethell & Zuber 2009) and the Hellas impact basin on Mars (e.g., Andrews-Hanna et al. 2008). At low impact angles (generally <30° from the horizontal), impactor breakup can result in ricochet and subsequent impacts of fragments downrange (Schultz et al. 2006), creating either a single elongated crater (e.g., Martian Orcus Patera) or an elliptical and circular crater pair (e.g., lunar Messier and Messier A; Schultz & Gault 1990). These workers suggest that the elongated crater in a pair is the site of the original impact, and the circular crater is created by the downrange ricochet of impactor fragments. Their findings also suggest that the width of elongated craters narrows downrange.

The information on oblique impact directionality given by impact crater/basin shape is important in terms of characterizing potential antipodal effects. Schultz & Crawford (2011) proposed substantial effects on the lunar surface antipodal to the initial contact site of the South Pole-Aitken (SPA) Basin-creating impactor. Via a combination of oblique impact experiments onto acrylic spheres and 3-dimensional hydrocode simulations, these workers demonstrated substantial physical disruption of the region roughly centered on the antipode to the point of first contact of the impactor with the target. They also noted that the center of the resulting topographic basin was offset downrange from the location of first contact. Based on these findings, Schultz & Crawford (2011) proposed an oblique impact scenario that produced a system of radial and circumferential tectonics covering much of the lunar near side, roughly antipodal to SPA. While recent magnetic field evidence for potential buried impactor material in the northwest SPA Basin rim (Wieczorek et al. 2012) suggests an impactor approach direction opposite to that favored by Schultz & Crawford (2011), the antipodal effects demonstrated by experiment and modeling are worth considering for Pluto.

We identify four locations within SP that are related to the impact process. The pink star in Figure 9 represents the potential first-contact point of an oblique impactor approaching from the NNW. The red triangle in Figure 9 marks the approximate center of the deep basin portion of SP (i.e., where the nitrogen ice is thick enough to support convection, McKinnon et al. 2016), identified as "the basin center." The yellow circle in Figure 9 marks the transition from the basin center to the southern portion of SP, where non-cellular plains suggest that the nitrogen ice is not convecting here and is shallower than in the basin center (McKinnon et al. 2016; White et al. 2017). The blue diamond in Figure 9 marks a potential first-contact point of an oblique impactor approaching from the SSE. Under a scenario where the approach of the SP-forming impactor was from the NNW (most closely resembling the scenario of Schultz & Crawford 2011), the initial site of impact is to the northwest of the basin center, placing its antipode beyond the southern limit of the imaged portion of the far side, in the dark southern latitudes.

If the incidence direction of the SP-forming impact were instead from the SSW, the point of first contact would more or less correspond to the teal diamond in Figure 10, with the excavated region widening until an ultimate circular-margined basin is produced. While the shape of SP in this incidence scenario does not correspond to the narrowing-downrange oblique impact scheme proposed for the irregular Orcus Patera crater on Mars (Schultz & Gault 1990), it does roughly match the "elliptical up range, circular downrange" pattern of the Messier and Messier A pair on the Moon, provided that the two phases of the impact were so close together in time and space that they essentially merged into one basin.

The SSW incidence scenario for SP may potentially account for the complex appearance of surface features located in a >1000 km diameter zone surrounding the first contact antipode (teal diamond in Figure 10), which includes the large-scale lineations of unit *dd* as well as the occurrences of the bladed terrain deposits (unit *btd*) and mottled, dark terrain (unit *mdt*) with especially angular and irregular planforms between 340°E and 10°E. The rather chaotic and intricate configuration of terrains within this zone may reflect disruption of Pluto's crust and surface topography in response to the Sputnik basin-forming impact, perhaps similar to the chaotic terrain seen antipodal to the Caloris basin on Mercury (Schultz & Gault 1975). Such disruption may include contributions from antipodal tectonics and deposition of ejecta at the antipode (Wieczorek & Zuber 2001).

In the above analysis, we consider the possibility that effects of the RTS and/or the SP basin-forming impact are manifested within the geology of the far side between 330°E and 20°E. However, the quality of the FS data is insufficient to distinguish which elements of the resolved geology may be due to one phenomenon or the other, or to determine if there are other geological processes besides these that are also affecting its appearance. More than any other portion of the far side, analysis of this region would benefit the most from improved resolution.

### Regarding Some Specific Far Side Features

We now discuss some other notable implications of the far side datasets and maps:

*Simonelli*. The largest identifiable impact crater on the mapped FS terrains is Simonelli, which at 250 km diameter rivals the largest identified impact crater on the near side hemisphere, Burney (~290 km diameter). Simonelli is located at the same latitude (12°N) as Elliot crater on the near side hemisphere. This places them at the northern limit of the permanent diurnal zone (13°N), which also marks the northern limit of the consistent surface layer of haze particles that defines Cthulhu Macula. Both Simonelli and Elliot have central peaks and high albedo floors, suggesting a common physical process of axially symmetric volatile deposition around the central peak. The NS impact craters Oort and Edgeworth are of broadly comparable size, but are located slightly south (~7.5° N) within Cthulhu Macula, and do not show high albedo floors. Whether this is related to their presence within Cthulhu Macula, or is due to some other cause, is unclear, but nonetheless demonstrate that the Elliot-Simonelli high albedo crater floors are not universal for craters of this size or even this latitude zone on Pluto.

*The Far Side Maculae and Their Relation to Bladed Terrain Deposits.* Far side mapping has revealed that the dark maculae, which in the near side hemisphere are represented by Krun Macula in the east and Cthulhu Macula in the west, are ubiquitous in their near equatorial latitude zone on the FS. Analysis of FS imaging, color, composition, and limb topography has also revealed that the bladed terrain deposits are much more widespread than is apparent based on observation of the near side hemisphere alone, where they are seen only at its eastern (Tartarus Dorsa) and western (label '4' in Figure 6C) margins. Exhibiting a strong methane spectral signature (e.g., Moore et al. 2018), these deposits represent the largest identified surface methane ice repository on Pluto, and can in some respects be considered a massive, high elevation counterpart to the low elevation, predominantly nitrogen ice deposit of Sputnik Planitia.

We also point out that the FS maculae are more fragmented than are the near side hemisphere maculae and may be intimately associated with the bladed terrain deposits that they border, as evidenced by the fact that the easternmost bladed terrain deposits are superposed on the dark, cratered plains of Cthulhu Macula. Whether this difference is due to volatile transport, altitude, or other effects, is unclear.

We hypothesize various scenarios for the formation of the FS maculae. The first is that they represent relatively low elevation bladed terrain deposits where, at some point in their history, sublimation and condensation of the surface methane ice in response to seasonal- and obliquity-driven climate change was interrupted for long enough to allow a layer of haze particles to accumulate (Grundy et al. 2018) and reach a sufficient thickness to prevent further volatile mobilization. Alternatively, the maculae may represent regions where the southern margin of the bladed terrain deposits underwent recession in response to secular climate change (as is hypothesized to have occurred at their northern margin), thereby exhuming a dark, water ice substrate. Being mostly located within the permanent diurnal zone, this inert water ice would accumulate a layer of haze particles, whereas water ice exhumed within the Arctic Circle oscillation range (Binzel et al. 2017) at the northern margin of the deposits would experience sufficient seasonal mobilization of volatiles to prevent accumulation of a consistent layer of haze particles. A third alternative is that the FS maculae are being exposed by volatile transport to higher latitudes, revealing a dark, underlying substrate.

*Additional Considerations Regarding Far Side Tectonics*: In addition to the RTS, Pluto possesses a global network of extensional faults (Stern et al. 2015; Moore et al. 2016). The orientation of these faults is not random—they follow coherent patterns over large swaths of Pluto—likely hinting at a common origin. At present, the favored model for this tectonic pattern due to Keane et al. (2016) who modeled the true polar wander of Pluto due to the formation of the Sputnik Planitia basin and glacier that dominates the anti-Charon hemisphere of Pluto. The combination of glacier loading, true polar wander, and global expansion (plausibly due to the freezing of a subsurface ocean, Hammond et al. 2016; Nimmo et al. 2016) produces a reasonable fit to the observed tectonic pattern on Pluto's near side.

If this hypothesis is correct, these same tectonic stresses should extend to Pluto's FS as well, as the stresses from true polar wander and global expansion are inherently global. The loading of sufficiently large features can also contribute to global stress patterns (e.g., Melosh 1980; Matsuyama et al. 2014). Figures 9 and 10 show the predicted tectonic pattern from Keane et al. (2016). Higher stress areas correspond to areas more prone to tectonic fracture.

The predicted stresses on Pluto's FS are slightly lower than on the near side owing to the decreased contribution of loading stresses. However, these stresses become similar to the stresses predicted near major faults on the near side of Pluto (e.g., Djanggawul, Virgil, and Sleipnir fossae) near the Sputnik Planitia antipode. Perhaps not coincidentally, this region is associated with alternating regions of north-south trending bright and dark regions (Vucub-Came, Hun-Came, and Meng-p'o Macula, and the far western edge of Cthulhu Regio), consistent with the predicted orientation of extensional faults in this model. It is thus conceivable that these geologic features may be related to such tectonic forces. Perhaps the tectonics partly or dominantly control the topography in this region, creating large-scale graben and horst, plausibly topped with bladed terrain.

However, it is important to note that while this model produces patterns that are suggestive of the surface markings in this antipodal region, this is not true everywhere. Pluto's FS is also associated with numerous linear markings (e.g., Chandrayaan Linea) further north. These lineations are not consistent with the predicted tectonic pattern—either suggesting that these features are not related to the tectonics, or that this tectonic model is incomplete. Unmodeled local tectonic stresses could significantly perturb the tectonic stresses on the FS. The Keane et al. (2016) tectonic model is also inconsistent with the putative N-S trending great ridge-trough system (RTS) identified in topography data (Schenk et al. 2018). One possible solution is that the RTS predates the global tectonic pattern observed today and that Pluto may have experienced multiple episodes of true polar wander and reorientation over its history.

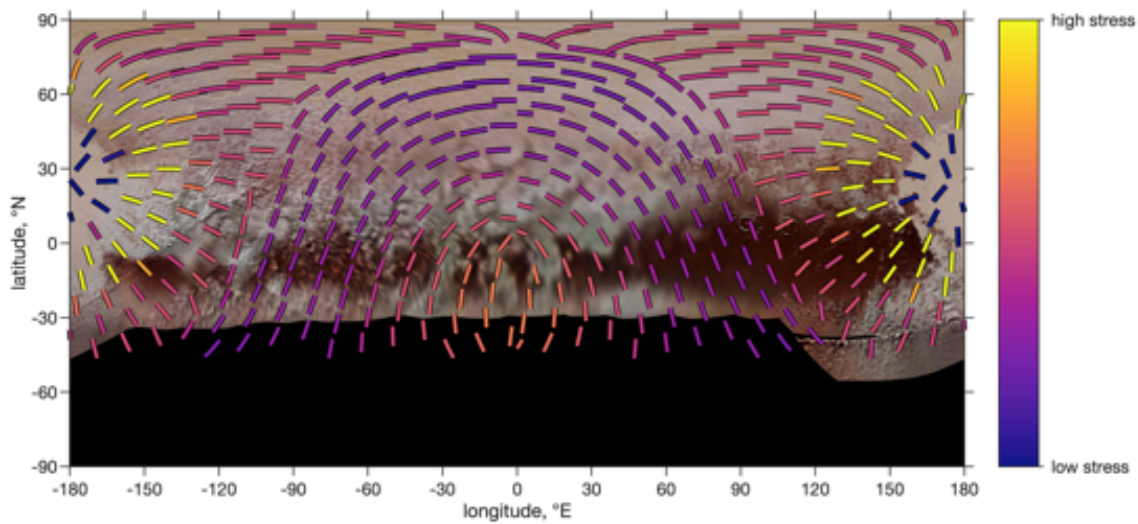

**Figure 11**: The predicted tectonic pattern on Pluto, due to loading of the Sputnik Planitia glacier, true polar wander, and global expansion (Keane et al. 2016). Line segments indicate the predicted orientation of extensional faults (i.e., graben), and the color of each line segment corresponds to the mean lithospheric stress. The exact magnitude of the stress depends on the extent of Pluto's global expansion, but are generally >1 MPa. The orientation of the predicted faults is not strongly sensitive to the magnitude of the stress. Proximal to Sputnik Planitia, loading stresses dominate, and produce faults that are quasi-radial to Sputnik Planitia. Distal to Sputnik Planitia, true polar wander stresses become important.

## Geological History

Figure 12 presents a correlation of most of the map units in Figure 5, which is a visual representation of how mapped geological units are oriented in space and time, relative to one another and established geologic time scales where known (Skinner et al. 2018).

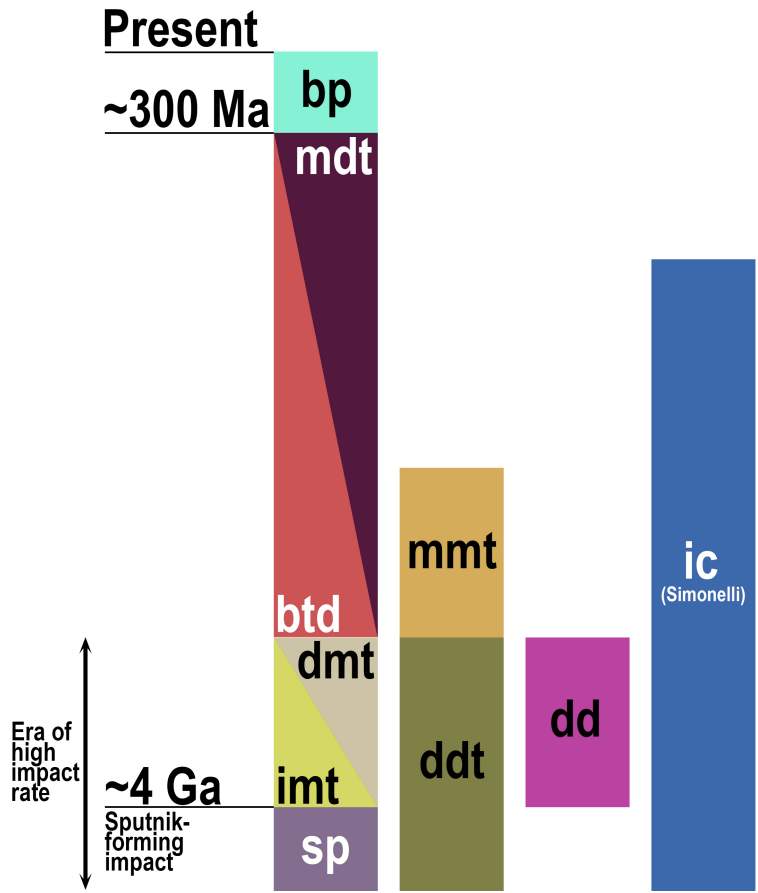

**Figure 12**. Correlation chart of geologic units mapped in Figure 5 (excepting *mit* and *ltm*, the natures of which are undetermined). Stratigraphic positions have been determined by crosscutting relations, crater size-frequency distributions of the units where expressed in the near side hemisphere, and topographic relief of the units where revealed in limb profiles. Unit boxes that are contiguous (*btd*/*mdt* and *imt*/*dmt*) are to indicate a close formative association. The top and bottom of each unit box corresponds to our assessment of the time scale of the formation and subsequent modification (if any) of each unit.

The earliest apparent FS unit is the smooth plains (unit *sp*) of western Cthulhu Macula, which may have formed approximately contemporaneously with the Sputnik basin-forming impact ≳4 Ga, or possibly even predating that. Consisting of inert $H_2O$-ice crust located within the permanent diurnal zone (Binzel et al. 2017), this terrain has experienced minimal modification by exogenic processes since it formed, save for (i) accumulating a thin layer of haze particles that have settled from the atmosphere (Grundy et al. 2018) and (ii) overprinting volatile transport.

The immediate post-Sputnik impact era (~4 Ga) likely witnessed emplacement of the bright, methane-rich mantle that covers Venera and Vega Terrae in the northwest of the near side hemisphere. The higher latitudes within the FS were also then mantled, forming the intermediate, mantled terrain (unit *imt*), which underwent extensive sublimation erosion in the mid-latitudes to form the dark, mantled terrain (unit *dmt*). This era was apparently characterized by a high impact rate, and the smooth plains and mantled terrains exhibit high crater counts (Singer et al. 2019).

Separate occurrences of the dark depressions (unit *dd*) may have different formation mechanisms, but it is speculated above that the regional-scale lineations formed by this unit between 330°E and 0°E may represent an extension of a giant north-south-aligned tectonic system that bisects the near side hemisphere (Schenk et al. 2018), or alternatively may indicate crustal disruption antipodal to the Sputnik basin-forming impact. Both hypotheses place the formation of the lineations at ~4 Ga, and if their floors are exposing dark water ice crust, then they must cut through (and therefore postdate) the surrounding intermediate, mantled terrain.

A thicker mantling deposit postdating that in Venera and Vega Terrae, then apparently covered Pioneer and Hayabusa Terrae to the northeast of Sputnik Planitia. This mantling, which displays fewer impact craters compared to the northwest (Singer et al. 2019), was likely emplaced as the impact rate was waning (i.e. <4 Ga), and may have been in part or whole sourced via gaseous emissions from large, deep pits in Pioneer Terra (Howard et al. 2017). The mottled, mantled terrain (unit *mmt*) within the far side appears to represent its eastern extent, and appears to be partly eroded via sublimation.

As noted above, the equatorial bladed terrain deposits (unit *btd*) form the largest repository of surface methane ice on Pluto, having originated during an episode of massive precipitation of atmospheric methane at high elevation within Pluto's low latitude diurnal zone, where averaged insolation is at a minimum (Binzel et al. 2017; Moore et al. 2018). Their timing of emplacement is difficult to estimate. They apparently postdate the smooth equatorial plains of Cthulhu Macula and the dark depressions and intermediate mantled terrain of the higher latitudes, but they must have formed early enough such that the deposition of the methane ice was not prevented by the accumulation of a consistent layer of dark haze particles.

Subsequent secular excursions in Pluto's climate seem to have partially eroded the bladed terrains, e.g., via sublimation, and may also have erased some impact craters. Where they occur within the near side hemisphere (Tartarus Dorsa), the bladed terrains are virtually un-cratered and exhibit a surface age of ~300 Ma (Moore et al. 2018), although on the far side, Simonelli crater (unit *ic*) may have impacted into the deposits. Recession of the deposits exhumed terrain that they previously superposed, including the dark, degraded terrain at their northern margin (unit *ddt*). The exact nature of the mottled, dark terrain (unit *mdt*) is uncertain, but its close geographical association with the bladed terrain deposits implies that the two units could be genetically related. The mottled, dark terrain may represent relatively low elevation deposits where mobilization of the surface methane ice was inactive for a period of time, or as noted above a substrate that was exhumed after recession of the southern margin of the bladed terrain deposits. In either scenario, the lack of volatile mobilization would lead to accumulation of a continuous blanket of dark haze particles on the surface (Grundy et al. 2018).

The bright plains (unit *bp*) are unequivocally the youngest terrain on the FS, consisting of predominantly nitrogen ice that has condensed on the floors of depressions amongst the bladed terrain deposits, and which is presently undergoing sublimation and glacial flow like similar deposits in East Tombaugh Regio in the near side hemisphere. The mottled, intermediate terrain (unit *mit*) and the light-toned material (unit *ltm*) are the least interpretable units on the FS, and accordingly the timing of their formation cannot yet be determined.

## Summation and Next Steps

Pluto's FS as seen by New Horizons displays a different terrain set than the near side hemisphere. Among the prominent expressions seen on the FS are: (i) evidence for the extensive global bladed terrain deposits not seen on the near side hemisphere; (ii) maculae that have different character than Cthulhu, which dominates the near side hemisphere's maculae; and (iii) an enigmatic complex of lineations that is crudely antipodal to and may relate to the origin of the SP forming event.

Future progress on far side geology, geophysics, and compositional studies would benefit tremendously from a Pluto orbiter. In particular, such an orbiter could address the key science questions including:

- ➢ What is the global distribution of Pluto's volatile units, and how does this relate to Pluto's climatic history?

- ➢ What is the nature of Pluto's great ridge-trough system (RTS), and is it truly global?

- ➢ How did Sputnik Planitia form, and did its formation play a significant role in shaping the far side geology of Pluto?

- ➢ Are the FS maculae being exhumed by volatile transport?

Since the arrival of any such orbiter is unfortunately at least two decades away, we must in the nearer term relay on advances that can be obtained from observatories with new capabilities on or near Earth, and laboratory and modeling efforts.

Most notably among those will be the new generation of 25-to-30 m telescopes coming on line in the 2020s. These devices will have diffraction limited resolutions ~10x better than Hubble Space Telescope, offering to obtain panchromatic, color, and even compositional spectroscopic maps of Pluto with resolutions of 30-50 km on a regular basis. The resolution of such maps will not surpass the New Horizons FS mapping discussed above, but will allow many more colors to be explored, and will also permit studies of surface albedo time variability as Pluto's complex orbital-obliquity seasons advance in the coming decades. In addition to yielding new knowledge about Pluto itself, the comparison of these dataset to similar datasets obtained on Triton and other Kuiper Belt dwarf planets will inform our understanding of the range of surface variation and variegation on such bodies as a class. Time-dependent Atacama Large Millimeter Array (ALMA) thermal mapping at crudely similar resolution will also be possible in the coming decades. The study of these small planets as a group with individually changing insolation and seasonal effects will better anchor volatile transport models used to understand them as a class.

## Acknowledgements

This work was supported, except as otherwise noted, by NASA's New Horizons mission. PJM was supported by NASA's New Frontiers Data Analysis Program grant 80NSSC18K1317, from which OLW also received partial support. This paper is dedicated to the future exploration of Pluto.

# References

Andrews-Hanna, J. C., M. T. Zuber, W. B. Banerdt (2008). The Borealis basin and the origin of the Martian crustal dichotomy. *Nature* **453**, 1212-1215.
doi: 10.1038/nature07011

Binzel R. P., Earle A. M., Buie M. W., et al. (2017). Climate zones on Pluto and Charon. *Icarus* **287**, 30-36.
doi: 10.1016/j.icarus.2016.07.023

Buie, M. W., Tholen, D. J, Horne, K. (1992). Albedo maps of Pluto and Charon: Initial mutual event results. *Icarus* **97**, 211-227.
doi: 10.1016/0019-1035(92)90129-U

Buie, M. W., Young, E. F., Binzel, R. P. (1997). "Surface Appearance of Pluto and Charon," in *Pluto and Charon* (S. A. Stern, D. J. Tholen, eds.), University of Arizona Press, Tucson, pp. 269-293.

Buie, M. W., Grundy, W. M., Young, E. F., Young, L. A., Stern, S. A. (2010). Pluto and Charon with the Hubble Space Telescope. II. Resolving Changes on Pluto's Surface and a Map for Charon. *AJ* **139**, 1128-1143.
doi: 10.1088/0004-6256/139/3/1128/meta

Buratti, B. J., Hofgartner, J. D., Hicks, M. D., et al. (2017). Global albedos of Pluto and Charon from LORRI New Horizons Observations. *Icarus* **287**, 207-217.
doi: 10.1016/j.icarus.2016.11.012

Cheng, A. F., Weaver, H. A., Conard, S. J., et al. (2008). Long Range Reconnaissance Imager on New Horizons. *Space Sci. Rev*. **140**, 189-215.
doi: 10.1007/s11214-007-9271-6

Earle, A. M., Binzel, R. P., Young, L. A., et al. (2017). Long-term surface temperature modeling of Pluto. *Icarus* **287**, 37-46.
doi: 10.1016/j.icarus.2016.09.036

Earle, A. M., Binzel, R. P., Young, L. A., et al. (2018). Albedo matters: Understanding runaway albedo variations on Pluto. *Icarus* **303**, 1-9.
doi: 10.1016/j.icarus.2017.12.015

Elbeshausen, D., Wünnemann, K., Collins, G. S. (2013). The transition from circular to elliptical impact craters. *J. Geophys. Res.* **118**, 2295-2309.
doi: 10.1002/2013JE004477

Garrick-Bethell, I., and Zuber, M. T. (2009). Elliptical Structure of the Lunar South Pole-Aitken Basin. *Icarus* **204**, 399.
doi: 10.1016/j.icarus.2009.05.032

Gladstone, G. R., Stern, S. A., Ennico, K., et al. (2016). The atmosphere of Pluto as observed by New Horizons. *Science* **351**, aad8866.
doi: 10.1126/science.aad8866

Grundy W. M., Binzel R. P., Buratti B. J., et al. (2016). Surface compositions across Pluto and Charon. *Science* **351**, aad9189.
doi: 10.1126/science.aad9189

Grundy W. M., Bertrand T., Binzel R. P., et al. (2018). Pluto's haze as a surface material. *Icarus* **314**, 232-245.
doi: 10 .1016/j.icarus.2018.05.019

Hammond, N. P., and Barr, A. C. (2014). Global resurfacing of Uranus's moon Miranda by convection, *Geology* **42**, 931–934.
doi: 10.1130/G36124.1

Hammond, N. P., Barr, A. C., Parmentier, E. M. (2016). Recent tectonic activity on Pluto driven by phase changes in the ice shell. *Geophys. Res. Lett.* **43**, 6775–6782.
doi: 10.1002/2016GL069220

Howard, A. D., Moore, J. M., White, O. L., et al. (2017). Pluto: Pits and mantles on uplands north and east of Sputnik Planitia. *Icarus* **293**, 218-230.
doi: 10.1016/j.icarus.2017.02.027

Johnson, B. C., Bowling, T. J., Trowbridge, A. J., Freed A. M. (2016). Formation of the Sputnik Planum basin and the thickness of Pluto's subsurface ocean. *Geophys. Res. Lett.* **43**, 10,068-10,077.
doi: 10.1002/2016GL070694

Keane, J. T., Matsuyama, I., Kamata, S., Steckloff, J. K. (2016). Reorientation and faulting of Pluto due to volatile loading within Sputnik Planitia. *Nature* **540**, 90-93.
doi: 10.1038/nature20120

Matsuyama, I., Nimmo, F., Mitrovica, J. X. (2014). Planetary reorientation. *Annu. Rev. Earth Planet. Sci.* **42**, 605–634.
doi: 10.1146/annurev-earth-060313-054724

McGovern, P. J., White, O. L., Schenk, P. M. (2019). Tectonism across Pluto: Mapping and Interpretations. *Pluto System After New Horizons*, Laurel, MD, Abstract #7063.

McKinnon, W. B., Nimmo, F., Wong, T., et al. (2016). Convection in a volatile nitrogen-ice-rich layer drives Pluto's geological vigour. *Nature* **534**, 82-85.
doi: 10.1038/nature18289

Melosh, H. J. (1980). Tectonic patterns on a reoriented planet: Mars. *Icarus* **44**, 745-751.
doi: 10.1016/0019-1035(80)90141-4

Moore, J. M., McKinnon, W. B., Spencer, J. R., et al. (2016). The geology of Pluto and Charon through the eyes of New Horizons. *Science* **351**, 1284-1293.doi: 10.1126/science.aad7055

Moore, J. M., Howard, A. D., Umurhan, O. M., et al. (2018). Bladed Terrain on Pluto: Possible origins and evolution. *Icarus* **300**, 129-144.
doi: 10.1016/j.icarus.2017.08.031

Moores, J. E., Smith, C. L., Toigo, A. D., Guzewich S. D. (2017). Penitentes as the origin of the bladed terrain of Tartarus Dorsa on Pluto. *Nature* **541**, 188-190.
doi: 10.1038/nature20779

Nimmo, F., Hamilton, D. P., McKinnon, W. B., et al. (2016). Reorientation of Sputnik Planitia implies a subsurface ocean on Pluto. *Nature* **540**, 94-96.
doi: 10.1038/nature20148

Nimmo, F., Umurhan, O., Lisse, C. M., et al. (2017). Mean radius and shape of Pluto and Charon from New Horizons images. *Icarus* **287**, 12-29.
doi: 10.1016/j.icarus.2016.06.027

Pappalardo, R. T., Reynolds, S. J., Greeley R. (1997). Extensional tilt blocks on Miranda: Evidence for an upwelling origin of Arden Corona. *J. Geophys. Res.* **102**, 13,369-13,379.
doi: 10.1029 /97JE00802

Protopapa, S., Grundy, W. M., Reuter, D. C., et al. (2017). Pluto's global surface composition through pixel-by-pixel Hapke modeling of New Horizons Ralph/LEISA data. *Icarus* **287**, 218-228.
doi: 10.1016/j.icarus.2016.11.028

Reuter, D. C., Stern, S. A., Scherrer, J., et al. (2008). Ralph: A visible/infrared imager for the New Horizons Pluto-Kuiper Belt mission. *Space Sci. Rev*. **140**, 129-154.
doi: 10.1007/s11214-008-9375-7

Rubincam, D. P. (2003). Polar wander on Triton and Pluto due to volatile migration. *Icarus* **163**, 469–478.
doi: 10.1016/S0019-1035(03)00080-0

Schenk, P. M., Beyer, R. A., McKinnon, W. B., et al. (2018). Basins, fractures and volcanoes: Global cartography and topography of Pluto from New Horizons. *Icarus* **314**, 400-433.
doi: 10.1016/j.icarus.2018.06.008

Schmitt, B., Philippe, S., Grundy, W. M., et al. (2017). Physical state and distribution of materials at the surface of Pluto from New Horizons LEISA imaging spectrometer. *Icarus* **287**, 229-260.
doi: 10.1016/j.icarus.2016.12.025

Schultz, P. H., and Gault, D. E. (1975). Seismic effects from major basin formations on the moon and mercury, *The Moon* **12**, 159-177.
doi: 10.1007/BF00577875

Schultz, P. H., and D. E. Gault (1990). "Prolonged global catastrophes from oblique impacts," in *Global Catastrophes in Earth History: an Interdisciplinary Conference on Impacts, Volcanism, and Mass Mortality* (V. L. Sharpton, P. D. Ward, eds.), Boulder, Colorado. *Geol. Soc. Am. Spec. Pap.* 247, p. 239.
doi: 10.1130/SPE247-p239

Schultz, P. H., Sugita S., Eberhardy, C. A., Ernst C. M. (2006). The role of ricochet impacts on impact vaporization. *Int. J. Impact. Eng.,* **33**, 771-780.
doi: 10.1016/j.ijimpeng.2006.09.005

Schultz, P. H. and Crawford, D. A. (2011). "Origin of nearside structural and geochemical anomalies on the Moon," in *Recent Advances and Current Research Issues in Lunar Stratigraphy* (W. A. Ambrose, D. A. Williams, eds.), *Geol. Soc. Am. Spec. Pap.* 477, pp. 141-159.
doi: 10.1130/2011.2477(07)

Singer K. N., McKinnon W. B., Gladman B., et al. (2019). Impact craters on Pluto and Charon indicate a deficit of small Kuiper belt objects. *Science* **363**, 955-959.
doi: 10.1126/science.aap8628

Skinner, J. A., Huff, A. E., Fortezzo, et al. (2018). Planetary Geologic Mapping Protocol-2018. USGS, Flagstaff, AZ.
https://astropedia.astrogeology.usgs.gov/alfresco/d/d/workspace/SpacesStore/01e32dcd-3072-4ac2-8e41-7cc5029bd2cf/PGM_Protocol_March2018.pdf

Stern, S. A., Bagenal, F., Ennico, K., et al. (2015). The Pluto system: Initial results from its exploration by *New Horizons*. *Science* **350**, aad1815.
doi: 10.1126/science.aad1815

Stern S. A., Grundy, W. M., McKinnon, W. B., Weaver, H. A., Young, L. A. (2018). The Pluto system after New Horizons. *ARAA*, **56**, 357-392.

doi: 10.1146/annurev-astro-081817-051935

Stofan, E. R., Hamilton, V. E., Janes, D. M., Smrekar, S. E. (1997). Coronae on Venus: Morphology and Origin," in Venus II: Geology, Geophysics, Atmosphere, and Solar Wind Environment (S. W. Bougher, D. M. Hunten, R. J. Phillips, eds.), University of Arizona Press, Tucson, pp. 931-965.

Trowbridge, A. J., Melosh, H. J., Steckloff, J. K., Freed, A. M. (2016). Vigorous convection as the explanation for Pluto's polygonal terrain. *Nature* **534**, 79-81.
doi: 10.1038/nature18016

White, O. L., Moore, J. M., McKinnon, W. B., et al. (2017). Geological mapping of Sputnik Planitia on Pluto. *Icarus* **287**, 261-286.
doi: 10.1016/j.icarus.2017.01.011

Wieczorek, M. A., and Zuber M. T. (2001). A Serenitatis origin for the Imbrian grooves and South Pole-Aitken thorium anomaly. *J. Geophys. Res.* **106**, 27,853-27,864.
doi: 10.1029/2000JE001384

Wieczorek, M. A., Weiss, B. P., Stewart S. T. (2012). An Impactor Origin for Lunar Magnetic Anomalies. *Science* **335**, 1212-1215.
doi: 10.1126/science.1214773

Young, E. F., and Binzel, R. P. (1993) Comparative Mapping of Pluto's Sub-Charon Hemisphere: Three Least Squares Models Based on Mutual Event Lightcurves, *Icarus* **102**, 134-149.
doi: 10.1006/icar.1993.1038

Young, E. F., Galdamez, K., Buie, M. W., Binzel, R. P., and Tholen, D. J. (1999). Mapping the Variegated Surface of Pluto. *AJ* **117**, 1063-1076.
doi: 10.1086/300722

Young, E. F., Binzel, R. P., and Crane, K. (2001). A Two-Color Map of Pluto's Sub-Charon Hemisphere. *AJ* **121**, 552-561.
doi: 10.1086/318008